\documentclass[10pt, conference, compsocconf]{IEEEtran}
\IEEEoverridecommandlockouts
\usepackage{cite}
\ifx\pdfoutput\undefined
	\usepackage{graphicx}
\else
	\usepackage[pdftex]{graphicx}
\fi
\graphicspath{{figures/}}
\usepackage{amsmath}
\usepackage{listings}
\usepackage{amsfonts}

\usepackage{xcolor}

\usepackage{amsfonts}
\usepackage{url}

\newcommand{\mm}{\textrm{M}}
\newcommand{\flux}{\textrm{F}}
\newcommand{\fluxn}[1]{\textrm{F}^{#1}}
\newcommand{\dofs}{\textrm{Q}}
\newcommand{\ncp}{\textrm{B}}
\newcommand{\ncpn}[1]{\textrm{B}^{#1}}

\newcommand{\ps}{\delta_{x_0}}
\newcommand{\proj}{\textrm{P}}
\newcommand{\numflux}{\mathcal{F}^*}
\newcommand{\dudx}{\textrm{D}}
\newcommand{\timeaverage}{\bar{q}_{\mathbf{l}s}(t_n)}

\newlength{\plotWidth}
\lstnewenvironment{pseudocode}[1][] 
{   
    \lstset{ 
        mathescape=true,
        frame=none,
        basicstyle=\footnotesize, 
        keywordstyle=\color{black}\bfseries\em,
        morekeywords={,for} 
        numbers=left,
        xleftmargin=.04\textwidth,
        escapechar=?,
        #1 
    }
}
{}
\definecolor{code_indent}{HTML}{000000}

\begin{document}

\title{Vectorization and Minimization of Memory Footprint for Linear High-Order Discontinuous Galerkin Schemes}

\author{
\IEEEauthorblockN{Jean-Matthieu Gallard, Leonhard Rannabauer, Anne Reinarz, Michael Bader}
\IEEEauthorblockA{Department of Informatics, Technical University of Munich\\
Email: \{gallard,leonhard.rannabauer,reinarz,bader\}@in.tum.de} 
}
\maketitle


\begin{abstract}

We present a sequence of optimizations to the performance-critical compute kernels of the high-order discontinuous Galerkin solver of the hyperbolic PDE engine ExaHyPE -- successively tackling bottlenecks due to SIMD operations, cache hierarchies and restrictions in the software design.

Starting from a generic scalar implementation of the numerical scheme, our first optimized variant applies state-of-the-art optimization techniques by vectorizing loops, improving the data layout and using Loop-over-GEMM to perform tensor contractions via highly optimized matrix multiplication functions provided by the LIBXSMM library. We show that memory stalls due to a memory footprint exceeding our L2 cache size hindered the vectorization gains. We therefore introduce a new kernel that applies a sum factorization approach to reduce the kernel's memory footprint and improve its cache locality. With the L2 cache bottleneck removed, we were able to exploit additional vectorization opportunities, by introducing a hybrid Array-of-Structure-of-Array data layout that solves the data layout conflict between matrix multiplications kernels and the point-wise functions to implement PDE-specific terms. 

With this last kernel, evaluated in a benchmark simulation at high polynomial order, only 2\% of the floating point operations are still performed using scalar instructions and 22.5\% of the available performance is achieved.

\end{abstract}


\section{Introduction}

Developing an exascale-ready solver framework for systems of partial differential equations (PDEs) is not a simple task. 
On the one hand, it requires fine-tuning of its performance-critical components with respect to both the user-written application 
and the target architecture on which the production code is expected to run. On the other hand, it must retain flexibility and usability. 
The EU Horizon 2020 project ExaHyPE (``An Exascale Hyperbolic PDE Engine'', \url{www.exahype.eu}) had just this goal: 
published as open source, ExaHyPE should allow 
medium-sized research teams to quickly realize extreme-scale simulations. It is intended as an \emph{engine} (as in ``game engine'') 
and while focusing on a dedicated numerical scheme and on a fixed mesh infrastructure, 
it provides flexibility in the PDE system to be solved~\cite{Reinarz:2019}.

ExaHyPE employs a high-order discontinuous Galerkin (DG) method combined with ADER (Arbitrary high-order DERivative) time-stepping first proposed in \cite{Titarev:2002}. 
A-posteriori Finite-Volume-based limiting addresses the problem of instabilities that may occur for non-linear setups \cite{Zanotti}.
The ADER method is an explicit one-step predictor-corrector scheme. The solution is first computed element-locally and then corrected using contributions from neighboring elements. 
A cache-aware, communication-avoiding ADER-DG realization \cite{Charrier} allows high-order accuracy in time with just a single (amortized) mesh traversal, which leads to high arithmetic intensity. 
This high arithmetic intensity leads to advantages in performance and time-to-solution compared to Runge-Kutta-DG (RK-DG) approaches (cf.~\cite{Dumbser:2018}).    
Previous performance studies \cite{Charrier:studies:2019} also showed, however, that these advantages can be impeded by a large memory footprint for the element-local predictor step. 
ADER-DG hence faces slightly different optimization challenges than the more widespread RK-DG methods.  

For flexibility and modularity, ExaHyPE isolates the performance-critical components of the ADER-DG scheme into compute kernels. 
Multiple variants of each kernel exist, allowing the user to adapt the scheme to a given application's numerical requirements -- 
for example, choosing between a scheme for a linear or a non-linear PDE system. 
Furthermore, code generation utilities are used before compilation to produce kernels that are tailored toward both the given application and the target architecture, 
as specified by the user \cite{gallard2019roleoriented}.

ExaHyPE's performance is heavily dominated by the Space Time Predictor (STP), 
which computes an entirely element-local time extrapolation of the solution. 
For linear PDE systems, such as in the context of seismic simulations \cite{kenneth:curvilinear:2019}, the STP is computed via a Cauchy-Kowalewsky scheme, which requires tensor operations 
that imply calls to PDE-specific \emph{user functions}. 
This generates conflicting requirements on the ExaHyPE API: the data layout needed for optimization of the tensor operations and that needed for vectorization of the user functions differs (AoS vs.\ SoA).

%

In this paper, we show the gradual optimization of the linear STP kernel starting from the generic non-optimized algorithm.
As each further optimization step also requires more work by the user, each new variant is added as an opt-in feature. 
Combined with the code generation approach used in ExaHyPE this ensures continuing sustainability of the code.

We start with a brief overview of the engine, the ADER-DG scheme, kernels and a focus on the linear STP kernel's Cauchy-Kowalewsky scheme. 
In Sec.~\ref{sec:baseOpt} we justify the choice of data layout and discuss the optimization of the STP kernel 
using ExaHyPE's code generation utilities and the LIBXSMM library \cite{libxsmm}.
Next we discuss how memory stalls impede efficiency of these optimizations and how to reformulate the kernel's algorithm to reduce its memory footprint and thus memory stalls.
In Sec.~\ref{sec:splitckvect} we describe how this removed bottleneck allows to exploit further vectorization opportunities by using a hybrid data layout to solve the data layout conflict caused by the API requirements.
Finally, we evaluate and compare the performance of all STP kernel variants in various benchmarks, 
focusing on the Intel Skylake architecture.







\section{The ExaHyPE Engine}

\subsection{ADER-DG solver}
\label{sec:ader}

The ExaHyPE engine is designed to solve a wide class of systems of linear and non-linear hyperbolic PDEs. However, in this paper we focus on efficiently solving linear equations. In matrix form the equations are as follows 
\begin{equation}
  \dofs_t = \mm \left(\nabla \cdot \left( \flux \left(\dofs\right) \right) + \ncp \cdot \nabla \dofs \right)+ \ps\textrm{,}
  \label{exahype_gen:pde_lin}
\end{equation}
where $\dofs$ represents the time and space dependent physical quantities of the system. The temporal evolution of the quantities is defined by the conservative flux $\flux\left( \dofs \right) = \left(F_1  \dofs,F_2 \dofs,F_3 \dofs\right)$, as well as the non-conservative flux term $\ncp \cdot \nabla  \dofs = \left(B_1 \nabla_x \dofs,B_2 \nabla_y \dofs ,B_3 \nabla_z \dofs\right)$.%
\footnote{$\nabla \cdot \flux$ denotes divergence of $F$, $\nabla \dofs$ the spatial gradient of $\dofs$.}
$\mm$ models material properties. Point sources are modeled via $\ps$.


In a DG scheme the numerical solution of \eqref{exahype_gen:pde_lin} is represented by polynomials within each cell. We use a nodal basis given by the Lagrange polynomials with either Gauss-Legendre or Gauss-Lobatto interpolation points.
The hexahedral mesh used in ExaHyPE allows us to use a tensor product basis, i.e.
each basis function is composed of one-dimensional basis functions $\Phi_{\mathbf{k}}\left(x,y,z\right) =  \phi_{k_1}\left(x\right)\phi_{k_2}\left(y\right)\phi_{k_3}\left(z\right)$. 
The multiindex $\mathbf{k}=(k_1,k_2,k_3)$ refers to a specific nodal coordinate.
 The unknown state vector can now be approximated in each element by $\dofs(x,y,z,t) \approx \sum_{\mathbf{k}}\phi_{k_1}(x)\phi_{k_2}(y)\phi_{k_3}(z)q_{\mathbf{k}}(t)$.

For brevity we refer to \cite{Zanotti, Reinarz:2019} for a more detailed introduction of the semi-discrete scheme, which relies on a strong formulation of \eqref{exahype_gen:pde_lin} and
can be summarized as follows.
\begin{itemize}
\item Multiply \eqref{exahype_gen:pde_lin} with a test function from the same space as the ansatz function and integrate over each element.

\item Integration by parts (twice) of the flux terms.
After this step we are left with volume and face integrals. To compute the face integrals we introduce numerical fluxes $\numflux$.  Here we assume that $\numflux$ is linear in $\dofs$ and $\flux$, which simplifies the algorithm significantly.

\item  Project the equation from the mesh elements onto the reference unit cube.
\end{itemize}

To solve linear problems of the form \eqref{exahype_gen:pde_lin} for suitable initial and boundary conditions we implemented the Cauchy-Kowalewsky (CK) method in ExaHyPE.

After the discretization in space the computation of the required volume and face integrals reduces to the computation of cell-local tensor operations. To evaluate the necessary integrals we apply a quadrature rule based on the nodal points chosen for the Lagrange polynomials. 
We require $N$ nodal points per dimension to achieve $N$-th order convergence. 

In each cell we must compute the following matrices:
\begin{itemize}
\item the discrete mass matrix $\mm$ --
   this results in a diagonal mass matrix with the quadrature weights as entries, saving us the effort of inverting the mass matrix. 
\item the discrete derivative operator $\dudx$.
\item a projection operator $\proj$ --
 this projection is needed to compute the effect of the point source on each node. It projects the point source onto each nodal basis function.
\item $\fluxn{1},\fluxn{2},\fluxn{3}$: the physical flux in each direction
\item $\ncpn{1},\ncpn{2},\ncpn{3}$: the non-conservative flux in each direction
\end{itemize}

In the resulting scheme we compute two kernels. 
The first incorporates all integrals in the reference element, which we call the volume term. 
Each degree of freedom is given by a multiindex for the basis function and an index for the variable:
    \begin{align*}
    V_{\mathbf{k}r, \mathbf{l}s} q_{\mathbf{l}s} &= \mm_{sr} \bigl(
      \dudx_{l_1 k_1} \fluxn{1}_{sr}
        q_{k_1 l_2 l_3 r}
     +  \ncpn{1}_{sr}  \dudx_{l_1 k_1}
       q_{k_1 l_2 l_3 r}
    \\
    & + \dudx_{l_2 k_2} \fluxn{2}_{sr}
     q_{l_1 k_2 l_3 r}
     + \ncpn{2}_{sr} \dudx_{l_2k_2}
     q_{l_1 k_2 l_3 r}
     \\ 
    &  + \dudx_{l_3  k_3} \fluxn{3}_{sr} 
     q_{l_1 l_2 k_3 r} 
     + \ncpn{3}_{sr} \dudx_{l_3k_3}
     q_{l_1 l_2 k_3 r}
     \bigr) +  \proj_{\mathbf{l}} \ps 
    \end{align*}
%
The second kernel includes all integrals along the boundary of the reference element. 
We denote the neighborhood of a given cell with $\mathcal{N}$, and then sum over 
the adjacent faces $f$:
  \begin{align*}
    \sum_{f \in \mathcal{N}} \numflux\bigl(q^f(t)_{\mathbf{k} r},\hat{q}^f(t)_{\mathbf{k} p},\flux^f(t)_{\mathbf{k} r},\hat{\flux}^f(t)_{\mathbf{k} r}\bigr) \textrm{,}
  \end{align*}
where $q^f$ denotes the projected degrees of freedom of the face and $\hat{q}^f$  those of the neighboring element.
 
In this notation the time-integrated semi-discrete scheme can be written as:
\begin{align}
  \begin{split}
    q_{\mathbf{k} r}&(t_{n+1}) = q_{\mathbf{k} r}(t_n)
    +  V_{\mathbf{k}r,\mathbf{l}s} \int_{t_n}^{t_{n+1}} q_{\mathbf{l}s}(t) dt\\
    &-\sum_{f \in \mathcal{N}} \int_{t_n}^{t_{n+1}}\numflux(q^f(t)_{\mathbf{k}r},\hat{q}^f(t)_{\mathbf{k}r},\flux^f(t)_{\mathbf{k}r},\hat{\flux}^f(t)_{\mathbf{k}r})
    \label{exahypeGenSemi}
  \end{split}
\end{align}
To complete the discretization we need to discretize \eqref{exahypeGenSemi} in time. The underlying assumption in this step is that for a small enough timestep $\Delta t$ the volume operator is approximately equal to the temporal derivative:
\begin{align}
  \frac{\delta q_{\mathbf{l}s}(t_n)}{\delta t} \approx  V_{\mathbf{k}r\mathbf{l}s} q_{\mathbf{l}s}(t_n) \label{exahype_gen:time_aspt}
\end{align}
By evolving $q_{\mathbf{l}s}(t)$ in a Taylor series of order $N$ and using \eqref{exahype_gen:time_aspt} we can compute \eqref{exahypeGenSemi}:
\begin{align}
  \begin{split}
    \int_{t_n}^{t_{n+1}} q_{\mathbf{k}r}(t) dt \approx \sum_{o=0}^{N} \frac{(\Delta t)^{o+1}}{\left(o+1\right)!} V^o_{\mathbf{k}r,\mathbf{l}s} q_{\mathbf{l}s}(t_n)   =: \timeaverage \textrm{,}
  \end{split}\label{exahypeTimeAverage}
\end{align}
The summands of \eqref{exahypeTimeAverage} can be computed  iteratively.
$$p_{o \mathbf{k} r} := V^o_{\mathbf{k}r, \mathbf{l}s} q_{\mathbf{l}s}(t_n) = V_{\mathbf{k}r,\mathbf{l}s} p_{o-1,\mathbf{l}s}.$$
The vector $p_{i\mathbf{k}p}$ is called Space Time Predictor (STP) and its computation is the main focus of this work. It operates only on cell-local information and is considerably more expensive to compute than the corrector step. Since $\numflux$ is a linear operation, the semi-discrete scheme can be transformed to
\begin{align}
  \begin{split}
    q_{\mathbf{k}r}&(t_{n+1}) = q_{\mathbf{k}r}(t_n) + V_{\mathbf{k}r,\mathbf{l}s}\timeaverage + \\
    &\sum_{f \in \mathcal{N}} \numflux\bigl(\bar{q}^f(t_n)_{\mathbf{k}r},\hat{\bar{q}}^f(t_n)_{\mathbf{k}r},\bar{\flux}^f(t_n)_{\mathbf{k}r},\hat{\bar{\flux}}^f(t_n)_{\mathbf{k}r} \bigr)
  \end{split} 
\end{align}
We call this the corrector step.

\subsection{Kernels and Linear STP}
\label{sec:linearSTP}

In this paper we discuss various implementations of the STP kernel.
For each implementation the output must be the required input of the corrector step, i.e. $\bar{q}_{\mathbf{l}s}(t_n)$, $\bar{q}_{\mathbf{k}r}^f(t_n)$ and $\bar{\flux}^f(t_n)_{\mathbf{k}r}$.
The projection onto the face of an element is performed by a single matrix-matrix multiplication, leaving no room for optimization.
We will thus only present implementations to compute $\bar{q}_{\mathbf{l}s}(t_n)$ and $\bar{\flux}_{\mathbf{l}s}(t_n)$.

The generic implementation of the STP follows the mathematical formulation of \eqref{exahypeTimeAverage}.
We iterate over the derivatives of the Taylor series of \eqref{exahypeTimeAverage} and store the whole STP (in the array $p$) and its fluctuations (in the array $dF$). Using these we can compute the time averaged degrees of freedom (in the array \texttt{qavg}) and fluctuations (in the array \texttt{favg}). See Fig.~\ref{fig:pseudocode} for the entire pseudocode.
\begin{figure}
\begin{pseudocode}
/* Compute the Space Time Predictor */  
p[0,k] = q[k] +
  pointSource(t=t_n, coordinate = k)
for ( o = 0; o < N; o++ ) {
  for ( k = 0; k < N^3; k++ ) {
    for ( d = 0; d < 3; d++ ) {
      flux[o,d,k] $\gets$ computeF(p[o,k], dim=d)
    }
  }
  for ( d = 0; d < 3 ; d++ ) {
    for ( k = 0; k < N^3; k++ ) {
      dF[o,d,k] $\gets$
        derive(flux[o,d,*], coordinate=k, dim=d)
    }
  }
  for ( d = 0; d < 3; d++ ) {
    for ( k = 0; k < N^3; k++ ) {
      gradQ[o,d,k] $\gets$
        derive(p[o,d,*], coordinate=k, dim=d)
    }
  }
  for ( d = 0; d < 3; d++ ) {
    for ( k = 0; k < N^3; k++ ) {
      dF[o,d,k] $\gets$ dF[o,d,k] +
        computeNcp(gradQ[o,k], dim=d)
    }
  }
  for ( k = 0; k < N^3; k++ ) {
    p[o+1,k] $\gets$ p[o,k] +
      derive(pointSource(t=*, coordinate=k),
             dim=time, order=o)
    for( d=0; d < 3; d++ ){
      p[o+1,k] $\gets$ p[o+1,k] + dF[o,d,k]
    }
  }
}

/* Compute the time averaged values */  
for ( o = 0  o < N; o++ ) {
  for ( k = 0; k < N^3; k++ ) {
    qavg[k] $\gets$ qavg[k] +  p[o,k] * 
                         ( dt^{o+1} / (o+1)! )
  }
}
for ( d = 0; d < 3; d++ ) {
  for ( o = 0; o < N; o++ ) {
    for ( k = 0; k < N^3; k++ ) {
      favg[d,k] $\gets$ favg[d,k] + dF[o,d,k] *
                         ( dt^{o+1} / (o+1)! )
    }
  }
}
\end{pseudocode}
\caption{Pseudocode for the generic implementation of the STP.}
\label{fig:pseudocode}
\end{figure}

\subsection{Engine structure}

\begin{figure}[t]
\centering
 \includegraphics[width=0.8\columnwidth]{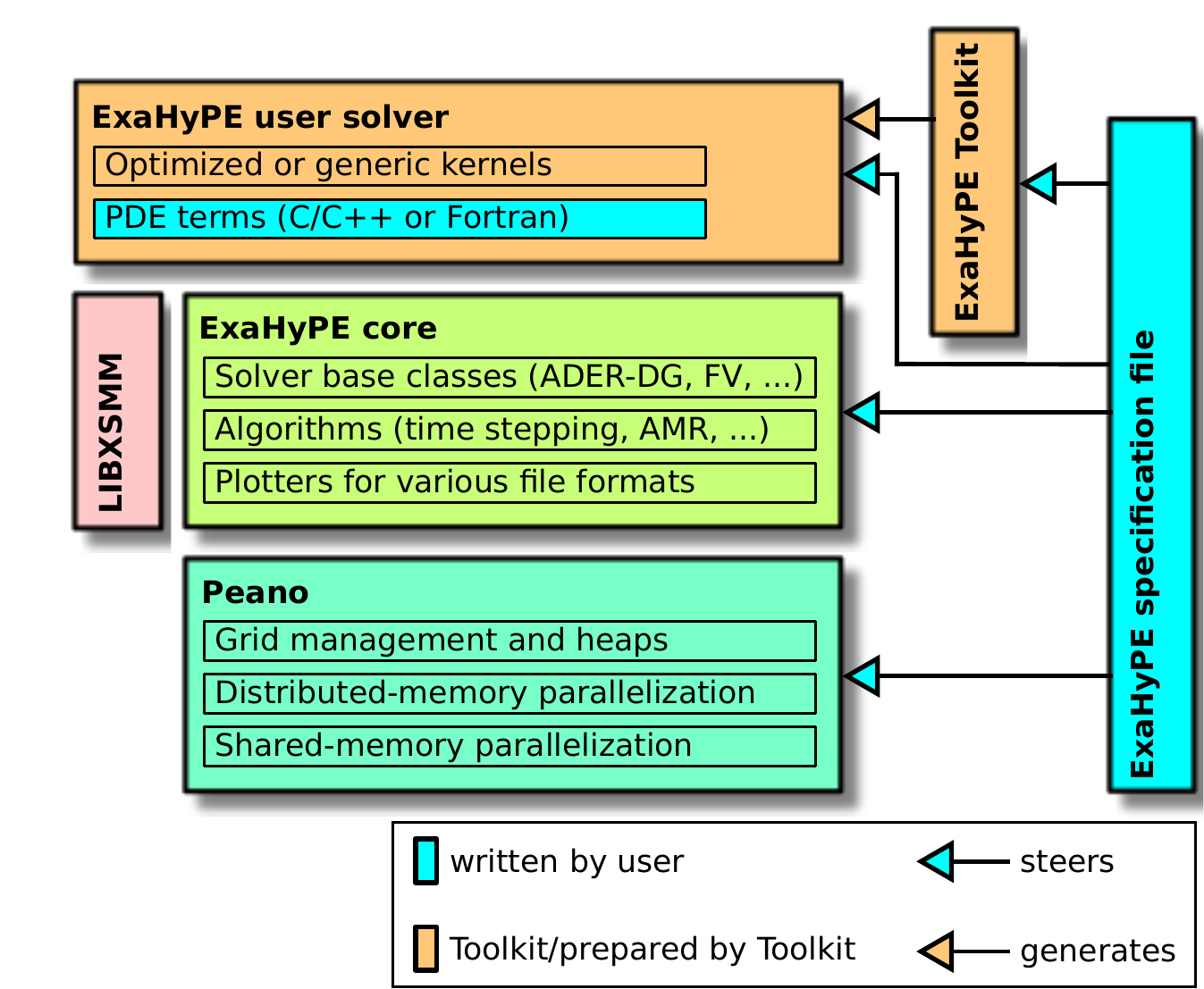}
 \caption{ExaHyPE's structure with kernels (taken from \cite{Reinarz:2019}).}
 \label{fig:arch}
\end{figure}

Fig.~\ref{fig:arch} shows the architecture of ExaHyPE, which follows a strict separation of concerns. 
It builds on the Peano framework \cite{Weinzierl:2019} (dark-green box),  
which provides dynamical adaptive mesh refinement on tree-structured Cartesian meshes together with shared- and distributed-memory parallelization. 

Application-specific contributions of users (i.e., developers of applications using the engine) are illustrated as cyan boxes.
To write an ExaHyPE application users provide a specification file, which is passed to the ExaHyPE \emph{Toolkit}. 
The Toolkit creates glue code, empty application-specific classes and (most important for this paper) core kernels that are tailored towards application and architecture (light-green box) \cite{gallard2019roleoriented}. 
Users need to complete the application-specific classes by providing PDE- and scenario-specific implementations (of flux functions, boundary conditions, etc.).
Note that these \emph{user functions} also depend on the requirements of the specification file and the kernel variant used.
For example, ExaHyPE's API by default relies on point-wise user functions that operate on a single quadrature node.
However some kernels, like the STP kernel variant described in Sec.~\ref{sec:splitckvect}, work on user functions that operate on vectorizable chunks. 

In the context of this paper it is also important to note the optional usage of LIBXSMM \cite{libxsmm} (red box). 
This library provides efficient routines for small matrix multiplications, which are used by the optimized ADER-DG kernels.

\subsection{Kernel Generator}

To generate application and architecture tailored kernels, the Toolkit uses a Kernel Generator submodule. The Kernel Generator is a Python 3 module that can be invoked manually or automatically by the Toolkit.
If invoked by the Toolkit, the optimized kernels are called by the generated glue code. 
The generic kernels used by default only provide minimal customizability toward the application and none toward the architecture.
The Kernel Generator, like the Toolkit, follows the Model-View-Controller (MVC) architectural pattern and uses the template engine Jinja2 to generate the C++ kernels.
This facilitates the introduction of new variants of a given kernel as a new template (View) in the module, and decouples the algorithmic optimizations from the low-level optimizations using Jinja2's template macro functionalities.
As architecture-specific optimizations are abstracted behind macros and template functionalities, these can easily be extended to support new architectures, e.g. by extending support for AVX-512 instruction sets.
A more detailed description of the Kernel Generator and Toolkit design can be found in \cite{gallard2019roleoriented}.

In the following sections, we introduce three variants of the generic STP kernel with increasing level of optimization. Each variant is implemented as a new template in the Kernel Generator 
and reuses existing optimization macros. 
As the later variants require an increasing amount of work from the end-user regarding the user functions, they are optional and can be enabled by a flag in the specification file.


\section{Optimized implementation of the linear STP kernel: vectorization and tensor operations}
\label{sec:baseOpt}

In this section we describe optimizations employed by the Kernel Generator to improve the STP kernel performance. These optimizations include 
SIMD vectorization and performing tensor contractions with \emph{Loop-over-GEMM} (LoG). The algorithm in this LoG variant of the STP kernel remains similar to  the generic variant in order to preserve the user API.
The use of slicing techniques to perform LoG by mapping tensor contraction to BLAS operations, in particular GEMM (matrix-matrix multiplication), has previously been explored by Di Napoli et al.~\cite{di2014towards} and Shi et al.~\cite{shi2016tensor}.


\subsection{SIMD and data layout conflict}

SIMD instructions require a unit-stride, i.e. the vectorized innermost loop must iterate over the fastest running index. 
For ExaHyPE, this causes conflicting data layout requirements. 
The kernels, in particular the STP kernel, perform the same operations for each quantity.
The operations can thus be vectorized in the quantity dimension, corresponding to an \emph{Array-of-Structure} (AoS) data layout.
In contrast, the user functions are evaluated pointwise for each quadrature node with the quantities as parameters. 
Vectorization can therefore only be performed by applying the user functions on multiple quadrature nodes at a time. 
This requires a \emph{Structure-of-Array} (SoA) data layout.
We expect the user functions to be less complex than the kernels and to be written by users who may be less familiar with code optimization. 
We therefore chose the AoS data layout for the ExaHyPE engine and thus prioritize optimization of the kernels.

Using the AoS data layout, kernel operations can be vectorized using compiler auto-vectorization with the corresponding pragmas or specialized code for critical operations.


Furthermore, to fully exploit the SIMD capabilities, all tensors and matrices are memory aligned and their leading dimensions are zero-padded to the next multiple of a SIMD vector length, such that each slice is also aligned in the slowest dimensions.
It should be noted, that the zero-padding will increase the number of floating point operations that need to be performed.
However, these come for free or even provide a speedup as a single vector instruction containing the zero-padding will replace one or more scalar instructions and also remove the need for masking or loop spilling. 

Since the alignment and padding-size depend on the target architecture supported SIMD instruction set, e.g. AVX-512 on Skylake, 
the Kernel Generator uses Jinja2's template variables and macros to abstract these optimizations in its templates.
These variables are calculated by the Kernel Generator Controller and are used by Jinja2 when rendering the macros and templates. 
The templates are rendered with the correct alignment and padding-size and future architectures can be added  by simply extending the macros' definitions, 
as was done when adding Knights Landing and Skylake (AVX-512) support from previous Haswell (AVX2) optimizations.

\subsection{Tensor products as Loop-over-GEMM with LIBXSMM}

The computation-heavy operations performed in the STP kernel reflect derivatives on tensors along one spatial dimension, 
as seen in the pseudocode of Fig.~\ref{fig:pseudocode}. These derivatives are computed as a matrix multiplication with the discrete derivative operator $\dudx$, 
as described in Sec.~\ref{sec:ader}.
This reduces the computation to tensor contraction operations.

For example, let $F_{\mathbf{k},s}$ be a 4-dimensional tensor with $\mathbf{k}=(k_1,k_2,k_3)$ corresponding to the $z$, $y$, and $x$ spatial dimensions respectively, and $s$ giving the quantity of the state-vector $q$ currently being computed.
Then the discrete derivatives $dF$ of $F$ along the $x$~dimension, take the form:
\[ \forall \mathbf{k},\, \forall s \colon~ dF_{\mathbf{k},s} = \sum_{l} \dudx_{k_3,l} F_{k_1,k_2,l,s}.\]
where $\dudx$ is specific to the quadrature nodes and  order.
If we first only consider the loops on $k_3$ and $s$, we need to compute the following intermediate result.
\[ \forall k_3,\, \forall s \colon~ dF'_{k_3,s} = \sum_{l} \dudx_{k_3,l} F'_{l,s},\]
which is a matrix multiplication with $F' = F_{k_1,k_2,:,:}$ the tensor~$F$'s matrix slice at a given $(k_1,k_2)$, and similarly $dF' = dF_{k_1,k_2,:,:}$.

Thus, we can perform the discrete derivative of $F$ more efficiently as a matrix multiplication on matrix slices of the tensors $F$ and $dF$ for fixed $k_1$ and $k_2$ values.
The computation of the tensor $dF$ can be reformulated using a sequence of BLAS $\texttt{MatMul}$ operations, as
\[\forall k_1,\,\forall k_2 \colon~ \texttt{MatMul}(F_{k_1,k_2,:,:},D,dF_{k_1,k_2,:,:}) \]

This reformulation of tensor contraction using batched matrix multiplications on matrix slices of the tensors is explored in more detail by Di Napoli et al.~\cite{di2014towards} and Shi et al.~\cite{shi2016tensor}, e.g.

Small matrix multiplications (matrix sizes are determined by the polynomial order $N$ of the ADER-DG scheme and the number $m$ of quantities in the PDE) are thus the performance-critical part of the STP kernel and need to be performed as efficiently as possible.
Exploitation of SIMD instructions demands a unit stride in the innermost loop of the matrix operations. 
As the quantity dimension (index $s$ in the example) is involved in all operations, using an AoS data layout implies that the quantity dimension should be the fastest running index of the tensors.

For highest-possible performance on Intel architectures, the Kernel Generator relies on the library LIBXSMM \cite{libxsmm}, 
which provides highly optimized assembly code for the required small dense matrix multiplications (\texttt{gemm} routines).

\begin{figure}[h]
\centering
\includegraphics[width=0.9\columnwidth]{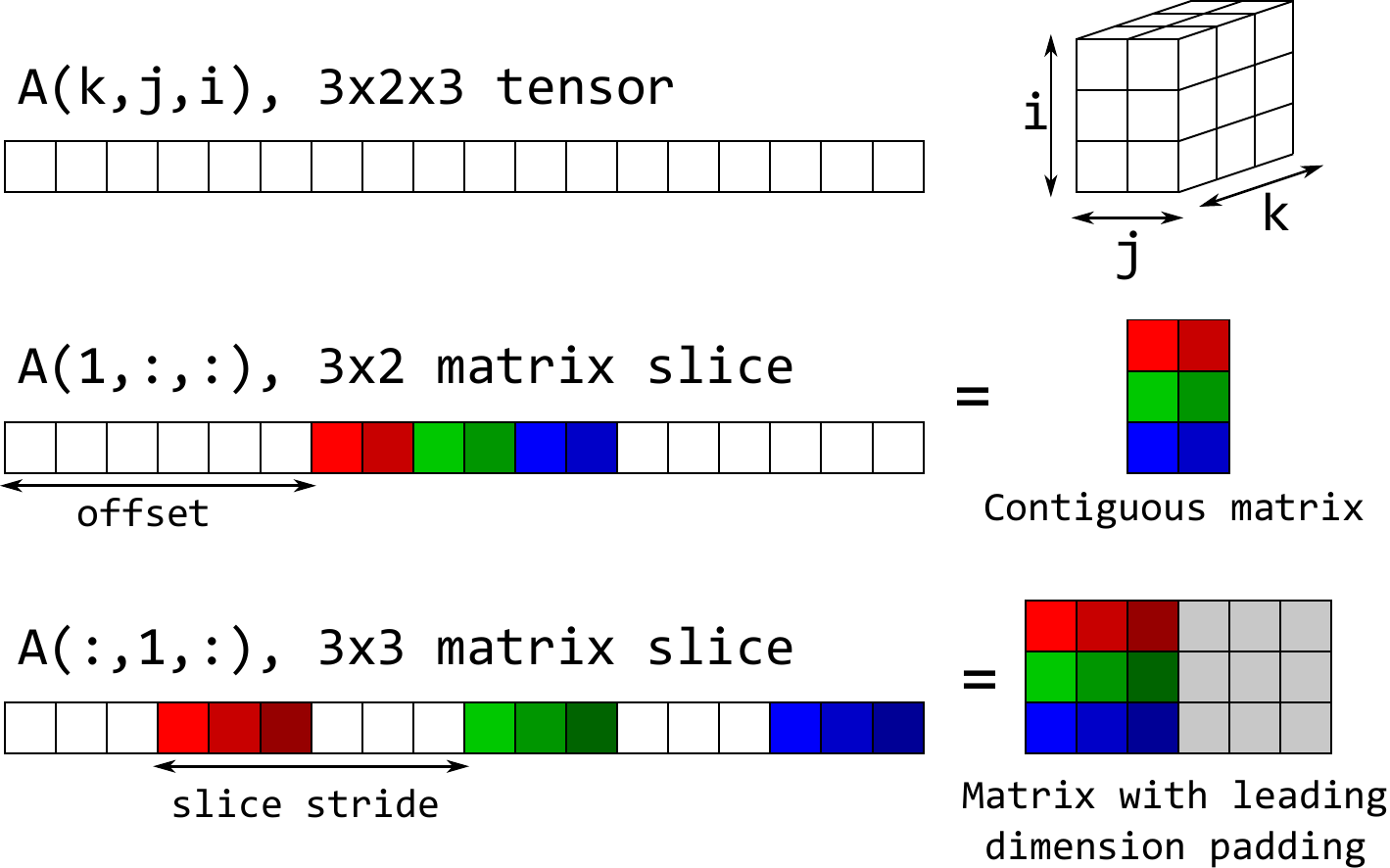}
\caption{Extracting matrix slices from a tensor. \label{fig:slice}}
\end{figure}

Fig.~\ref{fig:slice} illustrates how to extract matrix slices from the array storing a 3-dimensional tensor.
To extract a tensor matrix slice along the two fastest dimension (indexes $i$ and $j$) 
it is sufficient to specify an offset, as the matrix slice is a contiguous subarray of the array storing the tensor.
Doing so along other dimensions (indexes $i$ and $k$ in the figure) can be achieved by determining a \emph{slice stride}, if one of the dimensions features a unit stride (here $i$). 
The slice stride reflects the distance between the matrix rows that are stored in unit stride.
LIBXSMM allows us to define the leading dimension of a matrix with and without padding, e.g. to align matrix rows to cache line boundaries. 
We utilize these padding hints by interpreting the slice stride as the padded row size of the matrix leading dimension.
We thus efficiently restrict matrix operations to tensor matrix slices without requiring extra memory transfers.

Thus, all the tensor operations can be reformulated as a batch of matrix multiplications on subsequent slices in the different dimensions.
The zero-padding introduced earlier ensures that each slice is also aligned for optimal performance. 
We stress again that we rely on the quantity dimension being the leading dimension of the tensors and of the matrix slices. This requires an AoS data layout.

LIBXSMM is integrated into the Kernel Generator via a custom $\texttt{matmul}$ Jinja2 template macro as described in \cite{gallard2019roleoriented}.
This macro can also generate a generic triple-loop matrix multiplication, if LIBXSMM is not available for an architecture.
The use of a template macro facilitates integration of other libraries for small GEMM code, e.g.\ for other architectures. 
It also improves the readability of the template code by encapsulating the respective low-level optimizations.

\subsection{Further optimizations}

As a benefit of using code generation, compile time constants and known constant parameters are directly hard coded into the kernels -- as are the dimensionality of the problem and the location and exact name of the user functions. 
The latter, when combined with interprocedural optimizations (IPO) during compilation, allows inlining of the user functions in the kernels, even though these functions are virtual in the API.
In addition, frequently used matrices or combinations of matrices, such as cross products or the inverses of quadrature weights, can be precomputed by the Kernel Generator. They can then be directly hard coded, thus saving redundant operations.

\subsection{Intermediate performance results}

\begin{figure}
\centering
\includegraphics[width=\plotWidth]{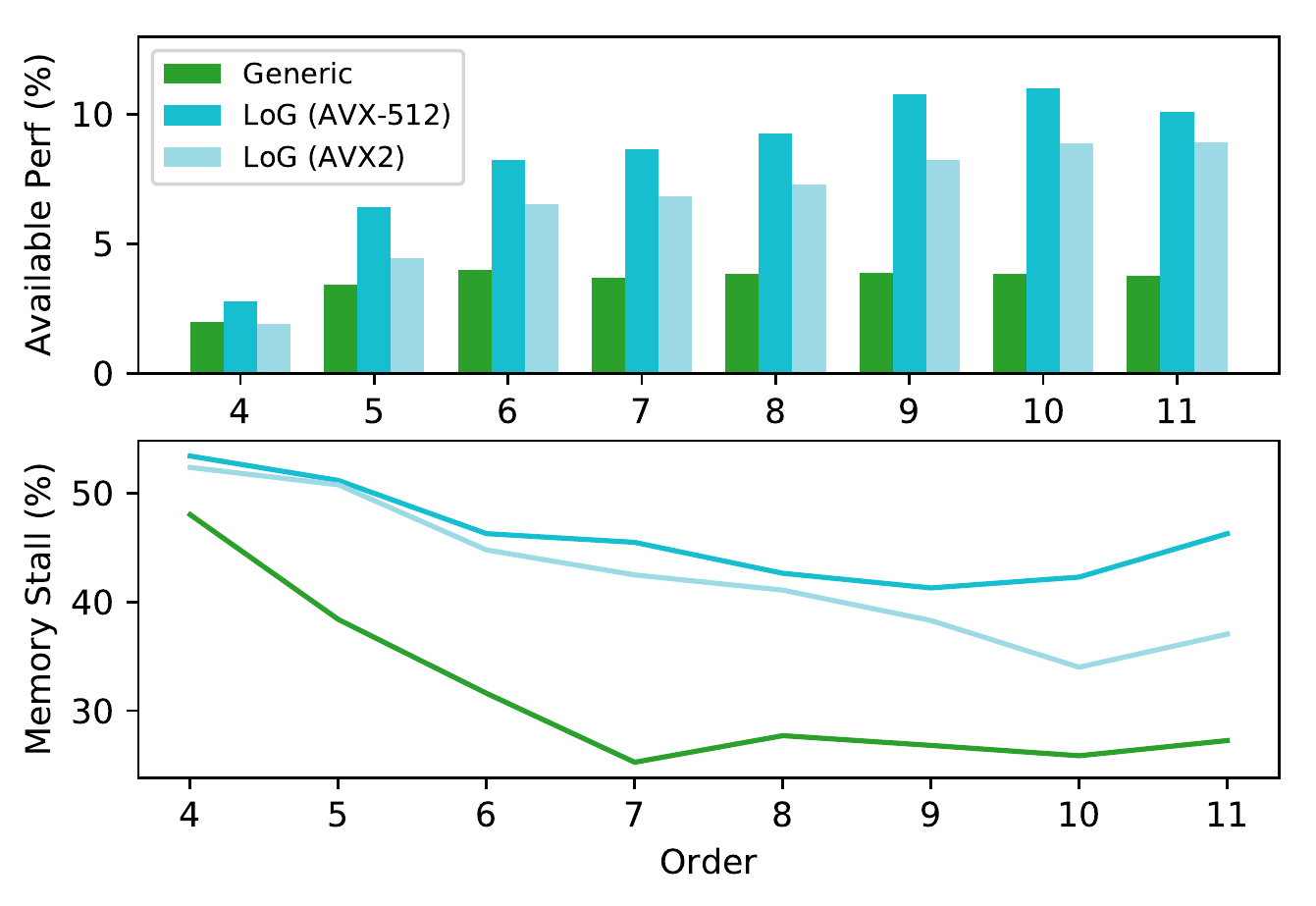}
\caption{Available performance reached and percentage of pipeline slots affected by memory stalls on the test benchmark (on Intel Skylake, see Sec.~\ref{sec:result}) 
with the generic (green), LoG for AVX-512 (turquoise), and LoG for AVX2 (light turquoise) kernels for order 4 to 11.
\label{fig:genvsoptPerf}}
\end{figure}

Fig.~\ref{fig:genvsoptPerf} compares the performance 
(as percentage of available performance\footnote{defined as the ratio of the average GFlops reached by the maximal amount of GFlops possible on this hardware as described in Sec.~\ref{sec:result}})
reached by the STP kernel in its generic variant vs.\ the LoG implementation that exploits the optimizations described in Sec.~\ref{sec:baseOpt}.
Benchmark and architecture (Intel Skylake) are described in detail in Sec.~\ref{sec:result}.
The LoG setup was tested with one setup optimized toward Skylake (AVX-512) and one toward the older Haswell architecture (AVX2) for comparison.

As expected the performance of the generic setup is quite low and quickly stagnates since only a fraction of the code can be auto-vectorized by the compiler. 
In contrast, more than 90\% of the floating point operations result from SIMD operations in the LoG AVX-512 setup (measured via Intel VTune Profiler, see Fig.~\ref{fig:instructionMix} in Sec.~\ref{sec:instructionMix}).
Neither of the LoG setups shows significant improvement at low order as the tensors are too small for the loop-vectorizations and matrix multiplications to be efficient. 
For higher order, they quickly improve to 8--11\% of the available performance.

However, the achieved performance does not reflect the potential expected from vectorization.
In fact, when comparing the LoG setup for Skylake and the setup optimized toward the previous Haswell architecture using AVX2, 
we obtain similar performance with a speedup of only 23--30\% obtained by going from AVX2 to AVX-512.
If the benchmarks were compute-bound, then a speedup closer to 100\% could be expected.
This discrepancy is explained by the significant amount of pipeline slots affected by memory stalls.
While they were expected, especially at low order when the arithmetic intensity of the ADER-DG scheme is lower, 
the percentage of pipeline slots impacted by memory stalls stays above 41\% in the LoG AVX-512 test and 34\% in the AVX2 setup. In the generic setup, however, the percentage goes down to around 27\%. At order 11 we observe a performance loss compared to order 9 and 10 for the LoG AVX-512 setup, which is due to the increase of memory stalls.


\section{SplitCK: Dimension splitted Cauchy-Kowalewsky scheme}
\label{sec:splitCK}

The LoG optimization of the STP kernel, as described in Sec.~\ref{sec:baseOpt}
did not lead to the desired performance improvement. 
In this section we show that this is due to memory stalls, most likely because the required data is not held in the fast cache levels. 
We therefore introduce a reformulation of the Cauchy-Kowalewsky scheme that reduces the memory footprint of the STP kernel.
Similar sum factorization approaches are used in other PDE solver frameworks, such as \cite{Kronbichler, Homolya, Muething, Schoeberl}.

\subsection{Motivation: LoG is bound by the capacity of the L2 cache}
\label{sec:splitckMotiv}

Performance benchmarks with the LoG kernel variant using Intel's VTune Amplifier display a high amount of pipeline slots impacted by memory stalls that plateau toward 40\% starting at order 6 instead of steadily decreasing as expected (see Fig.~\ref{fig:optvssplitck}). 
L2 cache overflow is to be expected around this order.
The benchmarks were performed on Intel Xeon Platinum 8174 CPUs, which have 1MB of L2 cache available per core.

For the generic scheme from Sec.~\ref{sec:linearSTP} the storage required for temporary arrays is $\mathcal{O}(N^{d+1}md)$, where $N$ is again the polynomial order of the ADER-DG scheme, 
$m$ the number of quantities in the PDE, and $d$ its dimension.
Thus, for a 3D medium-sized problem (e.g. $m=25$, $d=3$), the 1MB limit will be exceeded as soon as $N=6$ and temporary arrays will fall out of the L2 cache.

At high order the previous optimizations cannot be fully exploited, because the code stops being compute-bound and is instead dominated by memory stalls, mostly cache-related ones.
Therefore the algorithm described in Sec.~\ref{sec:linearSTP}, used by both the generic and LoG STP variants,
needs to be improved toward cache awareness and a reduced memory footprint, even at the cost of slightly more computations.

\subsection{Reformulation of the Cauchy-Kowalewsky scheme}

We reformulate the STP kernel's algorithm along the general paradigm to increase the number of memory accesses in lower level caches by reusing arrays as soon as possible.
Instead of computing and storing the fluxes for all dimensions at once, the tensor basis allows us to consider each dimension separately.
To decrease the overall memory footprint we do not keep the whole STP and its fluxes in memory, but directly add the contribution of each loop to the sum of the time averaged degrees of freedom and do not store the fluxes.
In the end we exploit the linearity of the scheme to recompute the time averaged fluxes from the time averaged degrees of freedom.

\begin{figure}
\begin{pseudocode}%[basicstyle=\footnotesize]
/* Compute qavg */
p[*]     $\gets$ q[*]
ptemp[*] $\gets$ 0
qavg[*]  $\gets$ dt * q[*]
for ( o = 0; o < N; o++ ) {
  for ( d = 0; d < 3; d++ ) {
    for ( k = 0; k < N^3; k++ ) {
      flux[k] $\gets$ computeF(p[k], dim=d)
    }
    for ( k = 0; k < N^3; k++ ) {
      ptemp[k] $\gets$
        derive(flux[*], coordinate=k, dim=d)
    }
    for ( k = 0; k < N^3; k++ ) {
      gradQ[k] $\gets$
        derive(p[*], coordinate=k, dim=d)
    }
    for ( k = 0; k < N^3; k++ ) {
      qtemp[k] $\gets$ qtemp[k] +
        computeNcp(gradQ[k], dim=d)
    }
  }
  for ( k = 0; k < N^3; N++ ) {
    qavg[k] $\gets$ qavg[k] + ptemp[k] *
                         ( dt^(o+1) / (o+1)! ) 
  }
  swap(p[k], ptemp[k])
}

/* Compute favg */
favg[*]  $\gets$ 0
for ( d = 0; d < 3; d++ ) {
  for ( k = 0; k < N^3; k++ ) {
    flux[k] $\gets$ computeF(p[k], dim=d)
  }
  for ( k = 0; k < N^3; k++ ) {
    favg[k] $\gets$ favg[k] +
      derive(flux[*], coordinate=k, dim=d)
  }
}
\end{pseudocode}
\caption{Pseudocode for the cache aware SplitCK scheme for the STP.}
\label{fig:SpliCK_code}
\end{figure}

The resulting scheme (outlined in Fig.~\ref{fig:SpliCK_code}) reduces the memory footprint of the largest tensors by a full time dimension by performing the time integration on-the-fly. A further reduction by a factor 3 is achieved by reusing the same tensors for all three spatial dimensions. 
The SplitCK memory footprint is thus $\mathcal{O}(N^{d}m)$, compared to $\mathcal{O}(N^{d+1}md)$ previously.
Cache locality is also taken into consideration.
However, the recomputation of the time averaged flux outside of the time loop adds the equivalent of almost one iteration  to the computation, a cost that becomes increasingly insignificant at higher order.

\subsection{Performance evaluation}

\begin{figure}
\centering
\includegraphics[width=\plotWidth]{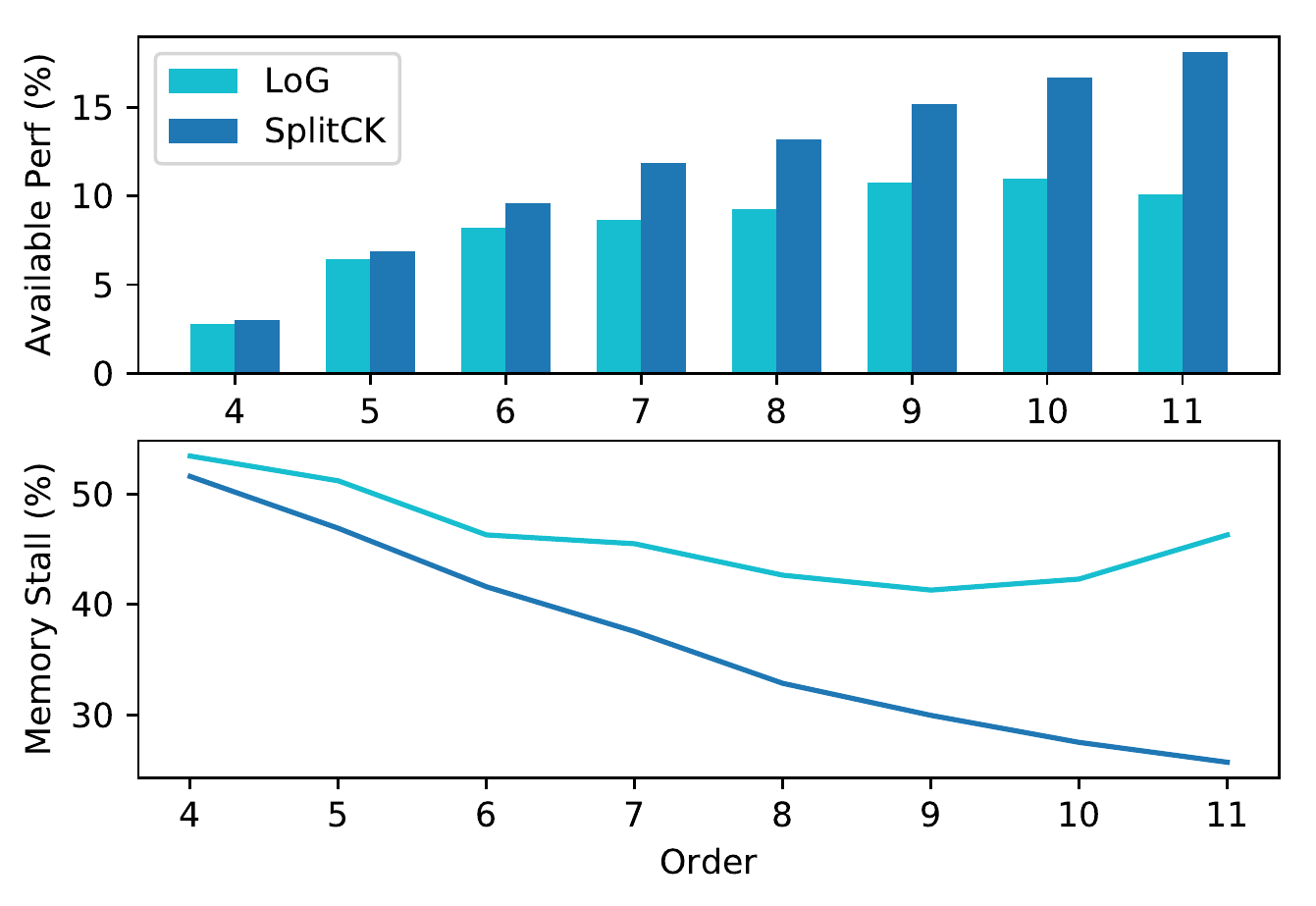}
\caption{Available performance reached and percentage of pipeline slots affected by memory stalls on the test benchmark (see Sec.~\ref{sec:result}) 
with the LoG variant (turquoise) and the SplitCK one (blue) for order 4 to 11.\label{fig:optvssplitck}}
\end{figure}

Fig.~\ref{fig:optvssplitck} shows that the new SplitCK scheme substantially reduces the memory stalls compared to the LoG scheme for all orders 
with a steady decrease as the order increases -- whereas the memory stalls for the LoG do not fall below 41\% and even increase after order 9.
The improvement in memory stalls is directly reflected by a performance that keeps growing with increasing order and comes closer to matching the expected speedup compared to the generic kernels.


\section{AoSoA data layout}
\label{sec:splitckvect}

Removing the L2 cache bottleneck via the SplitCK scheme not only leads to direct performance increases compared to the LoG scheme -- it enables further improvements by increasing the ratio of vectorized floating point operations. 
To achieve this, we address the AoS-vs.-SoA data layout conflict by introducing a hybrid data layout that can serve as AoS for the GEMM kernels and also as SoA for the user functions. Similar hybrid data layouts were also explored in the PyFR Navier–Stokes solver \cite{witherden2014pyfr} and are used by the YATeTo code generator \cite{Yateto} to optimize complex tensor operations.

\subsection{Motivation}

The data layout conflict between AoS and SoA and the choice of an AoS data layout means that vectorization opportunities in the user functions are lost as they are computed pointwise using scalar instructions.
To allow for SIMD instructions, they require inputs in an SoA data layout so that operations can be performed on a vector of individual quantities.

One way to get around this issue is to transpose the tensors on-the-fly to switch the data layout from AoS to SoA and back before and after calling the user functions.
This was tested in both linear and non-linear STP kernels for various applications.
It proved effective for complex non-linear scenarios, 
where the cost and complexity of the user functions were high enough that the performance gains of the vectorizations more than compensated for the cost of the transpositions.
However, the linear PDE systems in the targeted seismic applications have too simple (and inexpensive) user functions for such a solution to be effective, 
despite achieving the targeted high ratio of vectorized floating point operations.

By taking inspiration from the PyFR framework, where a similar conflict occurs \cite{witherden2014pyfr}, and from the optimization of tensor operations in YATeTo \cite{Yateto} 
(both are open source software), we implemented a hybrid Array-of-Structure-of-Array (AoSoA) data layout.
In this hybrid layout, the quantity dimension is put in between the spatial dimensions:
for a tensor $A$, instead of having the quantity dimension $s$ being the fastest, i.e. $A_{k,j,i,s}$ (in AoS layout), or the slowest, $A_{s,k,j,i}$ (in SoA layout), 
our AoSoA layout mixed it in between the spatial dimension, resulting in a $A_{k,j,s,i}$ hybrid data layout.
While this layout is slightly less intuitive to work with, it allows the kernels to keep working with a pseudo-AoS layout and to trivially extract SoA subarrays for the user functions.

To preserve alignment the fastest dimension is zero-padded as with the AoS data layout.
On AVX-512 architectures order~8 is a sweetspot with no padding required, whereas order 9 suffers from a particularly large padding overhead.

\subsection{Data layout in the kernel}

In the kernel, whenever tensor operations are performed on tensor's matrix slices, 
one of two cases can occur depending on which dimensions the slices are taken from: 

In the first case the slices are on the $x$~dimension, now the fastest running index. 
When compared to the previously used AoS data layout, the matrix slices of the AoSoA tensors are now transposed, as the quantity and $x$ dimensions where swapped in the tensors. 
Thus the matrix multiplications can simply be transposed too by using $C^T = (A \cdot B)^T = B^T \cdot A^T$. 
Performing the matrix multiplications in this case only requires to precompute the transpose of the second matrix $B$ (e.g. the discrete derivative operator matrix) 
and swap the tensor slice and $B^T$ in the \texttt{MatMul} operation.

In the second case the slices are on another dimension, i.e.\ on a slower running index than the quantity dimension. 
If both tensors have the same dimensions ordering and size, it is possible  -- when reformulating the tensor operations to Loop-over-GEMM -- to extract bigger slices by fusing multiple dimensions.
The code excerpt in Fig.~\ref{fig:fusedMatMul} shows a derivative along the $y$-dimension (index $j$) with the quantity and $x$-dimension fused (indexes $s$ and $i$), 
with an explicit matrix multiplication instead of a LIBXSMM GEMM.
Here, by fusing the quantity- and $x$-dimensions of the tensors when extracting the slices, it does not matter in which order they were in the initial tensor. This means that the same kind of matrix multiplications can be used as with the AoS data layout.
However, it forces minor API changes to ensure that the dimensions can be fused every time it is required.

\begin{figure}[h]
\begin{lstlisting}[basicstyle=\ttfamily\small]
for (int k = 0; k < 6; k++)
  // MatMul(Q, dudx, gradQ)
  // [j][is] slices, fuse i and s
  for (int j = 0; j < 6; j++)
    for (int l = 0; l < 6; l++)
      #pragma omp simd aligned([...])
      for (int is = 0; is < 72; is++)
        gradQ[k*432+j*72+is] += 
            Q[k*432+l*72+is] * dudx[j*8+l];
\end{lstlisting}
\caption{Tensor contraction expressed as Loop-over-GEMM with the quantity- and $x$-dimensions fused ($N=6$ and $m=12$ hard-coded). 
\label{fig:fusedMatMul}
}
\end{figure}

Vectorization of the STP kernel's operations can thus be preserved with the AoSoA data layout and LIBXSMM can still be used to optimize the \texttt{MatMul} operations.
However, the rest of the engine still expects an AoS data layout, thus the kernel inputs use the AoS data layout and are transposed to the AoSoA layout and the outputs are transposed back to AoS at the end of the STP kernel. 
The performance impact of these transpositions is minimal compared to the cost of the kernel itself.
In all cases this costs less performance-wise than using on-the-fly transposes between AoS and SoA at each user functions call.
Finally it could be avoided altogether by switching the whole engine to an AoSoA data layout. At the time of this paper this has not been done due to API compatibility issues with other parts of the engine.

\subsection{Vectorization of the user functions}


In the AoSoA STP kernel, instead of looping over all spatial dimensions to call a pointwise user function, 
the $x$ direction is now excluded from the loop and the vectorized user function is expected to be applied on the full $x$ dimension in one call.
When calling a user function the input subarrays of the AoSoA tensors are a full line of quadrature nodes, which corresponds to an SoA data layout within the subarray. Here the $x$ dimension is the fastest running index.

As the user functions are working on arrays in an SoA layout, they can easily be vectorized by looping over the fastest running index, replacing scalar operations by SIMD operations.
The $x$ dimension of the tensor is zero-padded to the next SIMD vector register size and the tensors are memory aligned, therefore each subarray is also memory aligned for maximal SIMD performance.

\begin{figure}[h]
\begin{lstlisting}[basicstyle=\ttfamily\small]
//scalar formulation of flux_x
void flux_x(double* Q, double* F) {
  F[0] = -(Q[0]+Q[3]+Q[4]);
  F[1] = -(Q[1]+Q[3]+Q[5]);
  F[2] = -(Q[2]+Q[4]+Q[5]);
}

//vectorized formulation of flux_x
void flux_x_vect(double* Q, double* F) {
  #pragma omp simd aligned(Q,F:ALIGNMENT)
  for(int i=0; i<VECTLENGTH; i++) {
    F[0*VECTSTRIDE+i] = -(Q[0*VECTSTRIDE+i]
       +Q[3*VECTSTRIDE+i]+Q[4*VECTSTRIDE+i]);
    F[1*VECTSTRIDE+i] = -(Q[1*VECTSTRIDE+i]
       +Q[3*VECTSTRIDE+i]+Q[5*VECTSTRIDE+i]);
    F[2*VECTSTRIDE+i] = -(Q[2*VECTSTRIDE+i]
       +Q[4*VECTSTRIDE+i]+Q[5*VECTSTRIDE+i]);
  }
}
\end{lstlisting}
\caption{Vectorization of a \texttt{flux\_x} user function. \label{fig:vectUserFunction}
}
\end{figure}

The code excerpt in Fig.\ref{fig:vectUserFunction} illustrates that in most applications, the scalar code can be adapted in three steps to enable the compiler auto-vectorization:
\begin{enumerate}
\item Enclose the code in a for loop to vectorize running over \texttt{VECTLENGTH}.
\item Add the new dimension to the arrays' indexes knowing that its size is \texttt{VECTSTRIDE}.
\item Use the correct pragma over the loop to enable auto-vectorization; here \texttt{omp simd} is used and an optional alignment specifier is given.
\end{enumerate}
\texttt{VECTLENGTH}, \texttt{VECTSTRIDE}, and \texttt{ALIGNMENT} are integer constants known at compile-time: 
\texttt{VECTLENGTH} is the size of the $x$ dimension without padding,
\texttt{VECTSTRIDE} the size with padding and thus the stride of the SoA data layout, and \texttt{ALIGNMENT} is the memory alignment adapted to the architecture.
In most cases, a simple compiler enabled auto-vectorization proved sufficient to achieve optimal performance when compared to user-written optimized code with Intel intrinsics for SIMD operations.

However, to achieve optimal performance in the user functions, special consideration is required regarding the zero-padding used to preserve alignment. 
The zero-padded vectors can cause issues in the user functions when zero is not a valid input value. This can lead to numerical errors such as division by zero. 
Thus, in most cases, the user function vectorization should only be performed on the non-padded dimension size.
But in doing so, masking or scalar loop spilling may be involved, decreasing performances.
Consequently, while vectorizing a user function is trivial, fully optimizing its performance requires careful consideration of potential numerical issues. 
For this reason these optimizations have been made available only as opt-in features.


\section{Results}
\label{sec:result}

We evaluate the performance of all the above-described kernel optimizations when simulating the elastic wave equations in first order formulation on curvilinear boundary-fitted meshes, as described in \cite{kenneth:curvilinear:2019}.
The equations are characterized by three quantities for particle velocity and six variables for the stress tensor. Three material parameters define density and the velocity of P- and S-waves.
To incorporate the geometry we store the transformation and its Jacobian in each vertex, adding a further nine parameters. Hence, we store $m=21$ quantities at each integration point. 
As a scenario we run the established LOH1 benchmark \cite{Day:LOH1}, with a curvilinear mesh to fit the elements to the material parameter interface.

All tests were performed on SuperMUC-NG at Leibniz Supercomputing Centre. 
SuperMUC-NG uses two Intel Xeon Platinum 8174 CPUs\footnote{\url{https://doku.lrz.de/display/PUBLIC/Details+of+Compute+Nodes}} per node, with 24 cores per socket running at 1.9GHz when using AVX-512.
Each core has two AVX-512 FMA units, thus the available performance per core is $1.9 \cdot 2 \cdot 2 \cdot 8 = 60.8$ double precision GFlops/s.
Note that the CPU base frequency is reduced by almost 30\%, from 2.7\,GHz to 1.9\,GHz, when using AVX-512 instructions. 
Thus, while these offer a theoretical speedup of a factor 8 in double precision FLOPs compared to scalar instructions, 
the actual speedup from vectorizing scalar operations is only a factor $\approx$5.6.
Tests were compiled using the Intel Compiler 19.0.5 with aggressive optimization flags\footnote{\texttt{-O3 -xCORE-AVX512 -restrict -fma -ip -ipo}}.

Multi-node parallelism of the ExaHyPE engine is handled by the Peano Framework \cite{Weinzierl:2019}, which uses a hybrid MPI+TBB (Intel Threading Building Blocks) approach. 
Due to NUMA effects, each MPI rank should be restricted to a single socket for optimal performance.
Furthermore Peano's task-based shared-memory parallelization begins to deteriorate beyond $\approx$10 cores, at least for our linear-PDE setup. 
We conclude that splitting each socket of 24 cores into 3 MPI ranks of 8~cores parallelized with TBB 
is the optimal layout used for large scale runs of the engine on SuperMUC-NG.

As we focus only on single node performance in this work, we therefore run all benchmarks on 8 cores parallelized with TBB on a single socket.

Performance is measured using Intel VTune Profiler 2019 on the full application, only excluding the engine initialization step, which is negligible in runtime for large scale runs.
The benchmarks thus measure end-to-end performance, with all kernels and engine overhead included -- though performance stays dominated by the STP kernel and its user functions.
The setups are named after the STP kernel variant used: generic (Sec.~\ref{sec:linearSTP}), LoG (Sec.~\ref{sec:baseOpt}), SplitCK (Sec.~\ref{sec:splitCK}) and AoSoA SplitCK (Sec.~\ref{sec:splitckvect}).

\subsection{Instruction mix}
\label{sec:instructionMix}

\begin{figure}
\centering
\includegraphics[width=\plotWidth]{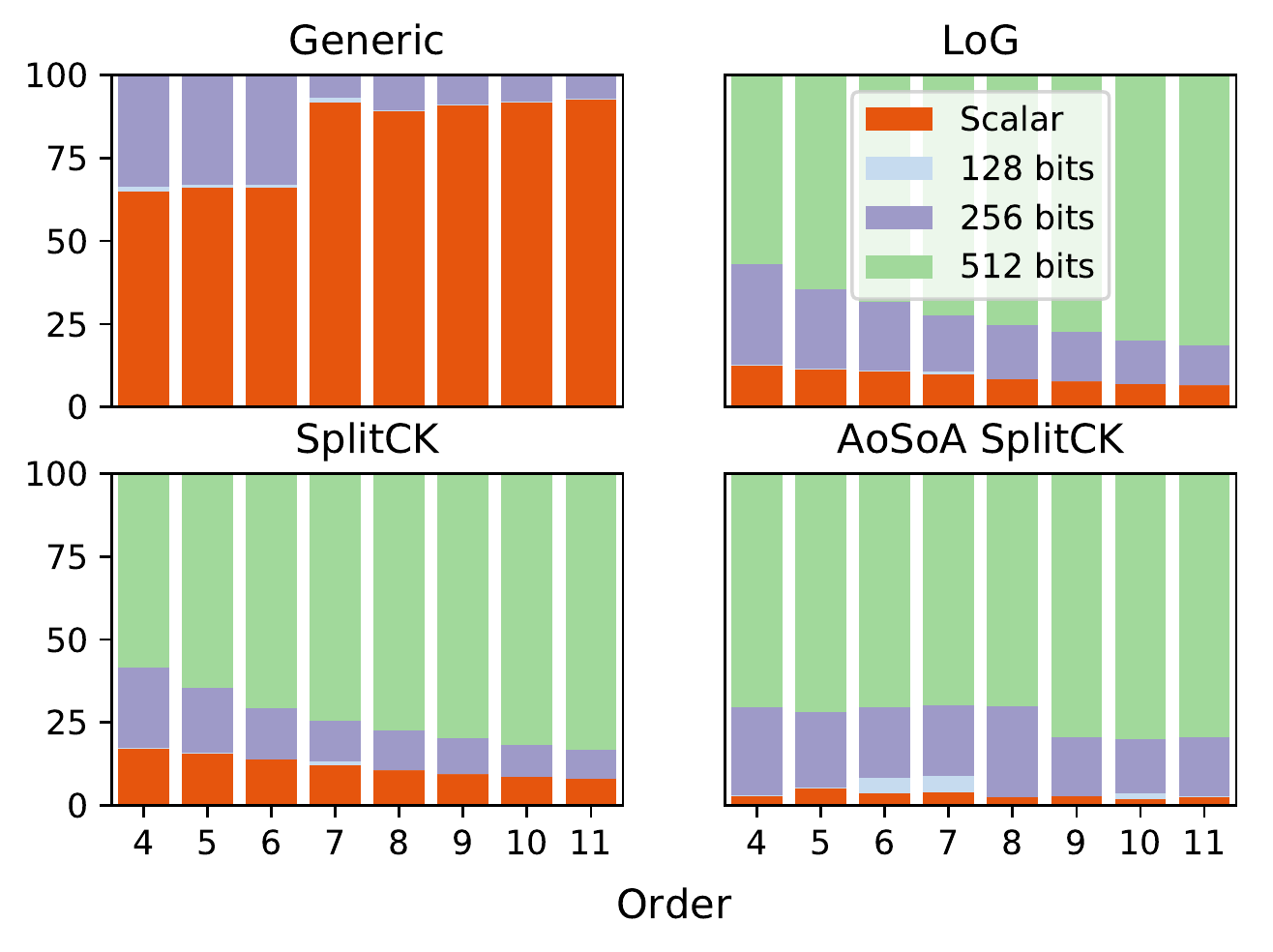}
\caption{Distribution of the packing size used to perform the floating point operations with the four kernel variants at orders $N= 4, \dots, 11$. \label{fig:instructionMix}}
\end{figure}

Fig.~\ref{fig:instructionMix} displays the SIMD instruction mix for the four optimization variants at increasing orders.
``Scalar'' corresponds to no vectorization; ``512 bits'' to SIMD packing with AVX-512 instructions.
As compiler auto-vectorization is used, 128- and 256-bit packing is also used following compiler heuristics.

For the generic setup, only a fraction of the FLOPs are packed by compiler auto-vectorization and most are computed using scalar instructions.
This changes with the LoG and SplitCK setups were over 80\% of the FLOPs are packed, either in 256- or 512-bit packed instructions.
This confirms that prioritizing the vectorization of the kernel was the right choice as it vectorizes most of the FLOPs.
However, still close to 10\% of the FLOPs, mostly coming from the user functions, are performed using scalar instructions.
Using the AoSoA SplitCK kernel, where the user functions are vectorized too, brings the amount of FLOPs performed with scalar instructions down to 2-4\%, close to full vectorization of the code.

\subsection{Available performance reached and memory stalls}

\begin{figure}
\centering
\includegraphics[width=\plotWidth]{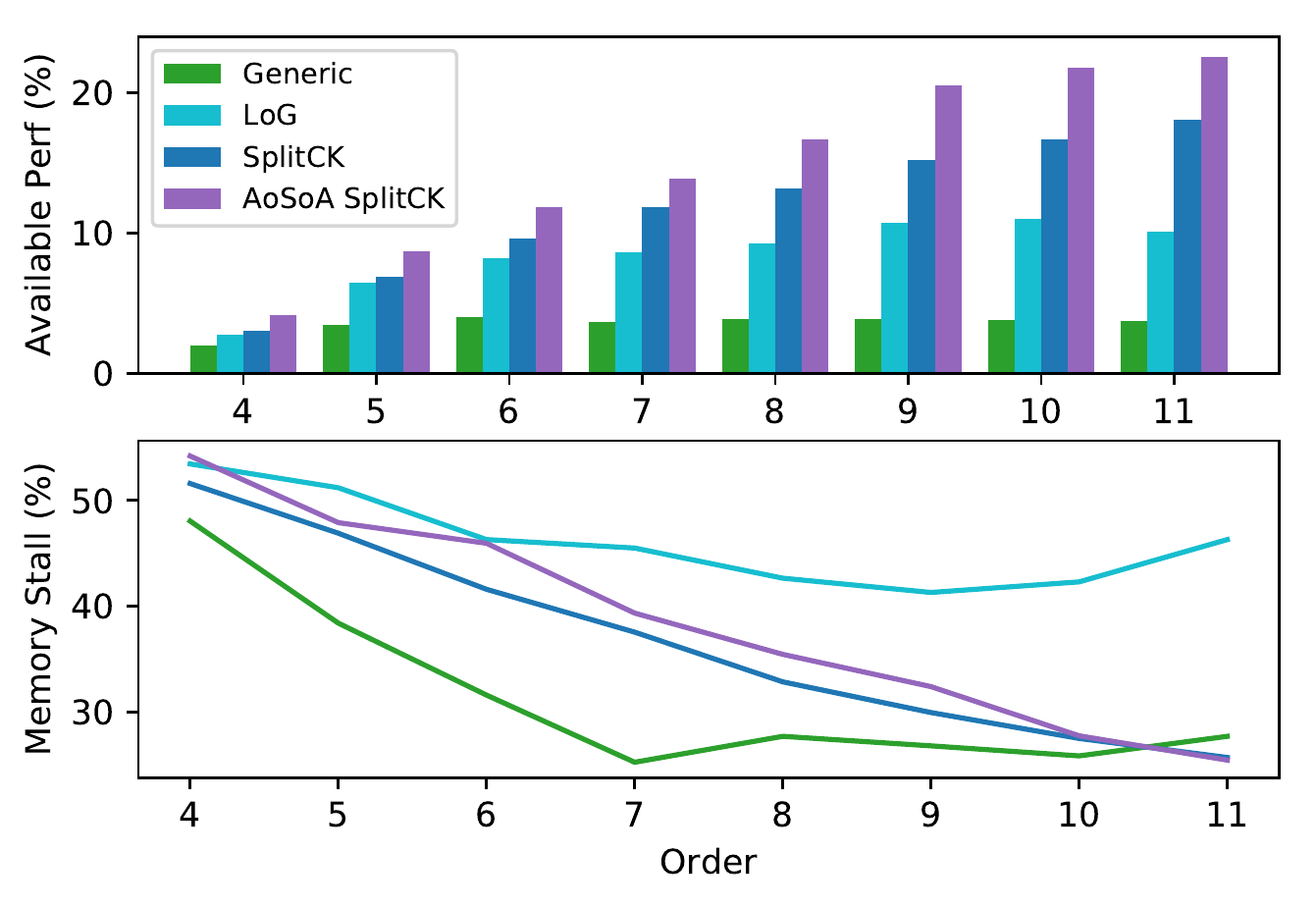}
\caption{Available performance reached and percentage of pipeline slots affected by memory stalls for all four kernel variants at orders $N=4,\dots,11$. \label{fig:perfAndMemStalls}}
\end{figure}

Fig.~\ref{fig:perfAndMemStalls} shows how much of the available performance was reached by each benchmark and plots the evolution of the ratio of pipeline slots affected by memory stalls.
As the ADER-DG scheme becomes more arithmetically intensive at higher order, the performance is expected to increase, while the memory stalls should steadily decrease for all kernel variants.
On the other hand, improving the compute speed of the STP kernels through vectorization and other optimizations increases the stress on memory and should result in an increased ratio of memory stalls compared to slower setups.

The generic kernels quickly reach a performance plateau around 3.8\%, which is consistent with expected performance from an almost scalar code on Skylake architectures \cite{Hammond, Kempf}.
The LoG setup is constrained by memory stall issues in its progress, starting from $N=6$, as discussed in Sec.~\ref{sec:splitckMotiv}.

Reducing the memory footprint of the STP kernel with the SplitCK scheme proved effective as the memory stalls ratio not only starts at a slightly lower value than in the LoG setup 
but also decreases faster (and steadily) with the order, even going below the generic setup for $N=11$.
Using the AoSoA SplitCK scheme to vectorize the user function does not impact the memory stalls significantly, 
slightly more are observed as this STP kernel variant is faster but the same trend as the default AoS SplitCK setup is preserved.
Thus, both SplitCK based setups increase in performances at higher orders with the setup using the AoSoA reaching 22.5\% at order $N=11$.
This is a speedup of a factor 6 compared to the generic benchmark at the same order. 
Taking into account the CPU base frequency reduction this is above the maximal speedup expected only from the AVX-512 vectorization of the algorithm. 
This is made possible by the reduced memory footprint of the new scheme.



\section{Conclusions}

%

To exploit the potential of modern CPU architectures such as Skylake, relying solely on the vectorization of the most critical operations is not sufficient to improve performance. As we have seen in ExaHyPE, not taking memory and caches into account can significantly impact the overall performance of the code.
Since the processor--memory performance gap is expected to further increase, 
this issue will become even more relevant on future architectures.
Additionally, the acceleration potential of AVX-512 implies that each missed vectorization opportunity can considerably impact performance. The AoSoA layout provided a considerable speedup despite only reducing the ratio of scalar FLOPs from around 10\% to around 3\% and in spite of adding the cost of computing transposes.

However, the modification of the STP kernel to reduce the memory footprint and improve vectorization also impacted the user API 
and increase the users' programming effort, if they want to use the SplitCK and hybrid layout scheme. The introduced hybrid AoSoA data layout is also less intuitive to use than the common AoS or SoA data layouts, which complicates development and maintenance of the code. 

Code generation proved to be a very valuable tool to solve these two issues.
For ExaHyPE users, the new variants, SplitCK and AoSoA SplitCK, are optional and can easily be introduced later by adapting an application developed for the simpler default user API.
For the engine developers, the template macros and high level abstractions available in the Kernel Generator greatly simplify the data layout change.


\section{Acknowledgements and Funding}

This project has received funding from the European Union's Horizon 2020 research and innovation programme under grant agreements No 823844 (project ChEESE, 
\url{www.cheese-coe.eu}) and No 671698 (ExaHyPE, \url{www.exahype.eu}).

We thank the Gauss Centre for Supercomputing e.V.\ (\url{www.gauss-centre.eu}) for providing computing time on the GCS supercomputers SuperMUC and SuperMUC-NG at Leibniz Supercomputing Centre (\url{www.lrz.de}).

We especially want to thank Carsten Uphoff for his support and advice on optimization of small tensor operations.

\bibliographystyle{IEEEtran}
\bibliography{IEEEabrv,references}

\begin{thebibliography}{10}
\providecommand{\url}[1]{#1}
\csname url@samestyle\endcsname
\providecommand{\newblock}{\relax}
\providecommand{\bibinfo}[2]{#2}
\providecommand{\BIBentrySTDinterwordspacing}{\spaceskip=0pt\relax}
\providecommand{\BIBentryALTinterwordstretchfactor}{4}
\providecommand{\BIBentryALTinterwordspacing}{\spaceskip=\fontdimen2\font plus
\BIBentryALTinterwordstretchfactor\fontdimen3\font minus
  \fontdimen4\font\relax}
\providecommand{\BIBforeignlanguage}[2]{{%
\expandafter\ifx\csname l@#1\endcsname\relax
\typeout{** WARNING: IEEEtran.bst: No hyphenation pattern has been}%
\typeout{** loaded for the language `#1'. Using the pattern for}%
\typeout{** the default language instead.}%
\else
\language=\csname l@#1\endcsname
\fi
#2}}
\providecommand{\BIBdecl}{\relax}
\BIBdecl

\bibitem{Reinarz:2019}
A.~{Reinarz}, D.~E. {Charrier}, M.~{Bader}, L.~{Bovard}, M.~{Dumbser},
  K.~{Duru}, F.~{Fambri}, A.-A. {Gabriel}, J.-M. {Gallard}, S.~{K{\"o}ppel},
  L.~Krenz, L.~Rannabauer, L.~Rezzolla, P.~Samfass, M.~Tavelli, and
  T.~Weinzierl, ``{ExaHyPE:} an engine for parallel dynamically adaptive
  simulations of wave problems,'' \emph{Comp. Phys. Comm}, accepted,
  arXiv:1905.07987.

\bibitem{Titarev:2002}
V.~A. Titarev and E.~F. Toro, ``{{ADER:} Arbitrary High Order {Godunov}
  Approach},'' \emph{J. Scient. Comput.}, vol.~17, no.~1, pp. 609--618, Dec
  2002.

\bibitem{Zanotti}
O.~Zanotti, F.~Fambri, M.~Dumbser, and A.~Hidalgo, ``Space–time adaptive
  {ADER} discontinuous {Galerkin} finite element schemes with a posteriori
  sub-cell finite volume limiting,'' \emph{Comp.\ Fluids}, vol. 118, pp.
  204--224, 2015.

\bibitem{Charrier}
D.~E. Charrier and T.~Weinzierl, ``Stop talking to me -- a
  communication-avoiding {ADER-DG} realisation,'' \emph{arXiv e-prints}, 2018,
  arXiv:1801.08682.

\bibitem{Dumbser:2018}
M.~Dumbser, F.~Fambri, M.~Tavelli, M.~Bader, and T.~Weinzierl, ``Efficient
  implementation of {ADER} {D}iscontinuous {G}alerkin schemes for a scalable
  hyperbolic {PDE} engine,'' \emph{Axioms}, vol.~7, no.~3, p.~63, 2018.

\bibitem{Charrier:studies:2019}
D.~E. Charrier, B.~Hazelwood, E.~Tutlyaeva, M.~Bader, M.~Dumbser,
  A.~Kudryavtsev, A.~Moskovskiy, and T.~Weinzierl, ``Studies on the energy and
  deep memory behaviour of a cache-oblivious, task-based hyperbolic {PDE}
  solver,'' \emph{Int. J. High Perf. Comp. App.}, vol.~33, no.~5, pp. 973--986,
  2019.

\bibitem{gallard2019roleoriented}
\BIBentryALTinterwordspacing
J.-M. Gallard, L.~Krenz, L.~Rannabauer, A.~Reinarz, and M.~Bader,
  ``Role-oriented code generation in an engine for solving hyperbolic {PDE}
  systems,'' in \emph{Workshop on Software Engineering for {HPC}-Enabled
  Research ({SE-HER 2019})}, SC19, Denver, 2019. [Online]. Available:
  \url{https://arxiv.org/abs/1911.06817}
\BIBentrySTDinterwordspacing

\bibitem{kenneth:curvilinear:2019}
K.~Duru, L.~Rannabauer, O.~K.~A. Ling, A.-A. Gabriel, H.~Igel, and M.~Bader,
  ``A stable discontinuous {Galerkin} method for linear elastodynamics in
  geometrically complex media using physics based numerical fluxes,''
  \emph{arXiv preprint arXiv:1907.02658}, 2019.

\bibitem{libxsmm}
A.~Heinecke, G.~Henry, M.~Hutchinson, and H.~Pabst, ``{LIBXSMM:} accelerating
  small matrix multiplications by runtime code generation,'' in \emph{{SC16:}
  Int. Conf. for HPC, Networking, Storage and Analysis}, 2016, pp. 981--991.

\bibitem{Weinzierl:2019}
T.~Weinzierl, ``The {Peano} software---parallel, automaton-based, dynamically
  adaptive grid traversals,'' \emph{ACM Trans. Math. Softw.}, vol.~45, no.~2,
  pp. 14:1--14:41, 2019.

\bibitem{di2014towards}
E.~Di~Napoli, D.~Fabregat-Traver, G.~Quintana-Ort{\'\i}, and P.~Bientinesi,
  ``Towards an efficient use of the {BLAS} library for multilinear tensor
  contractions,'' \emph{Appl. Math, Comput.}, vol. 235, pp. 454--468, 2014.

\bibitem{shi2016tensor}
Y.~Shi, U.~N. Niranjan, A.~Anandkumar, and C.~Cecka, ``Tensor contractions with
  extended {BLAS} kernels on {CPU} and {GPU},'' in \emph{2016 IEEE 23rd Int.
  Conf. on High Perf. Comp. (HiPC)}.\hskip 1em plus 0.5em minus 0.4em\relax
  IEEE, 2016, pp. 193--202.

\bibitem{Kronbichler}
M.~Kronbichler and K.~Kormann, ``A generic interface for parallel cell-based
  finite element operator application,'' \emph{Comp.\ Fluids}, vol.~63, pp.
  135--147, 2012.

\bibitem{Homolya}
M.~Homolya, R.~C. Kirby, and D.~A. Ham, ``Exposing and exploiting structure:
  optimal code generation for high-order finite element methods,'' \emph{arXiv
  e-prints}, 2017, arXiv:1711.02473.

\bibitem{Muething}
S.~M\"uthing, M.~Piatkowski, and P.~Bastian, ``High-performance implementation
  of matrix-free high-order discontinuous {Galerkin} methods,'' \emph{Int. J.
  High Perf. Comp. App.}, 2018.

\bibitem{Schoeberl}
J.~Sch\"oberl, A.~Arnold, J.~Erb, J.~M. Melenk, and T.~P. Wihler, ``C++11
  implementation of finite elements in ngsolve,'' \emph{Technical report},
  2017.

\bibitem{witherden2014pyfr}
F.~D. Witherden, A.~M. Farrington, and P.~E. Vincent, ``{PyFR:} an open source
  framework for solving advection--diffusion type problems on streaming
  architectures using the flux reconstruction approach,'' \emph{Comp. Phys.
  Comm.}, vol. 185, no.~11, pp. 3028--3040, 2014.

\bibitem{Yateto}
\BIBentryALTinterwordspacing
C.~Uphoff and M.~Bader, ``Yet another tensor toolbox for discontinuous
  {Galerkin} methods and other applications,'' \emph{ACM Trans. Math. Softw.},
  under review. [Online]. Available: \url{http://arxiv.org/abs/1903.11521}
\BIBentrySTDinterwordspacing

\bibitem{Day:LOH1}
S.~M. Day and C.~R. Bradley, ``Memory-efficient simulation of anelastic wave
  propagation.'' \emph{Seismol. Soc. Am., Bull.}, vol.~3, pp. 520--531.

\bibitem{Hammond}
S.~D. Hammond, C.~T. Vaughan, and C.~Hughes, ``Evaluating the {Intel Skylake
  Xeon} processor for {HPC} workloads,'' \emph{2018 Int. Conf. on High Perf.
  Comp. \& Sim. (HPCS)}, pp. 342--349, 2018.

\bibitem{Kempf}
D.~Kempf, R.~He\ss, S.~M\"uthing, and P.~Bastian, ``Automatic code generation
  for high-performance discontinuous {Galerkin} methods on modern
  architectures,'' 2018, arXiv:1812.08075.

\end{thebibliography}


\end{document}